\documentclass{amsart}
\usepackage{amsmath,amsfonts,amssymb}
\usepackage{mathrsfs,geometry}
\usepackage[T1]{fontenc}
\usepackage[latin1]{inputenc}
\usepackage{url}

\newcommand{\iy}{\ensuremath{\infty}}

\newcommand{\C}{\ensuremath{\mathbf{C}}}

\renewcommand{\P}{\ensuremath{\mathbf{P}}}
\newcommand{\e}{\varepsilon}

\renewcommand{\leq}{\leqslant}
\renewcommand{\geq}{\geqslant}
\renewcommand{\phi}{\varphi}
\newcommand{\Id}{\mathrm{Id}}

\newcommand{\noi}{\noindent}

\DeclareMathOperator{\tr}{Tr}

\newtheorem*{theo}{Theorem}

\newtheorem*{lemma}{Lemma}
\newtheorem*{rk}{Remark}

\newtheorem*{defn}{Definition}

\begin{document}

\author{Guillaume AUBRUN}
\title{A remark on the paper ``Randomizing quantum states: Constructions and applications''}

\maketitle

\begin{abstract}
The concept of $\e$-randomizing quantum channels has been introduced by Hayden, Leung, Shor and Winter  in connection with approximately encrypting quantum states. They proved using a discretization argument that sets of roughly $d \log d$ random unitary operators provide examples of such channels on $\C^d$. We show that a simple trick improves the efficiency of the argument and reduces the number of unitary operators to roughly $d$.
\end{abstract}

\bigskip

Since our argument is  a minor modification of the original proof, we systematically refer the reader to \cite{hlsw} for introduction, background and applications of the notion of randomizing states.

\bigskip

\textit{Notation.}
On the space $\mathcal{B}(\C^d)$ of $d \times d$ complex matrices we consider the trace class norm $\|\cdot\|_1$ and the operator norm $\|\cdot\|_\iy$. Let also $\mathcal{D}(\C^d)$ be the convex set of mixed states (=positive elements of $\mathcal{B}(\C^d)$ with trace 1). The extreme points of $\mathcal{D}(\C^d)$ are pure states. We denote by $C$ and $c$ absolute numeric constants.

\begin{defn}
A quantum channel (= completely positive trace-preserving linear map) $R : \mathcal{B}(\C^d) \to \mathcal{B}(\C^d)$ is said to be $\e$-randomizing if for every state $\phi \in \mathcal{D}(\C^d)$,
\[ \left\| R(\phi)-\frac{\Id}{d} \right\|_{\iy} \leq \frac{\e}{d}. \]
\end{defn}

\begin{theo}
Let $(U_i)_{1 \leq i \leq N}$ be independent random matrices Haar-distributed on the unitary group $\mathcal{U}(d)$. Let $R : \mathcal{\C}^d \to \mathcal{\C}^d$ be the quantum channel defined by
\[ R(\phi) = \frac{1}{N} \sum_{i=1}^N U_i \phi U_i^\dagger .\]
Assume that $0<\e<1$ and $N \geq Cd/\e^2 \cdot \log(1/\e)$. Then the channel $R$ is $\e$-randomizing with nonzero probability.
\end{theo}

\noi As often with random constructions, we actually prove that the conclusion holds true with \textit{large} probability. Let us quote two lemmas from \cite{hlsw}.

\begin{lemma}[Lemma II.3 in \cite{hlsw}]
Let $\phi,\psi$ be pure states on $\C^d$ and $(U_i)_{1 \leq i \leq N}$ as before. Then for every $0<\delta<1$,
\[ \P \left( \left| \frac{1}{N} \sum_{i=1}^N \tr (U_i\phi U_i^\dagger \psi) -\frac{1}{d} \right| \geq \frac{\delta}{d} \right) \leq 2 \exp (-c\delta^2 N)\]
\end{lemma}

\begin{lemma}[Lemma II.4 in \cite{hlsw}]
For $0<\delta<1$ there exists a set $\mathcal{M}$ of pure states on $\C^d$ with $|\mathcal{M}| \leq (5/\delta)^{2d}$, such that
for every pure state $\phi$ on $\C^d$, there exists $\phi_0 \in \mathcal{M}$ such that $\|\phi-\phi_0\|_1 \leq \delta$.
\end{lemma}

\noi \textit{Proof of the theorem.} 
Let $A$ be the (random) quantity
\[ A = \sup_{\phi,\psi \in \mathcal{D}(\C^d)} \left| \frac{1}{n} \sum_{i=1}^n \tr (U_i \phi U_i^\dagger \psi) - \frac{1}{d} \right|. \]
We must show that $\P(A \geq \frac{\e}{d}) < 1$. Let $B$ be the restricted supremum over the set $\mathcal{M}$
\[ B = \sup_{\phi_0,\psi_0 \in \mathcal{M}} \left| \frac{1}{n} \sum_{i=1}^n \tr (U_i \phi_0 U_i^\dagger \psi_0) - \frac{1}{d} \right|. \]
It follows from the lemmas that for $\delta$ to be determined later
\[ \P \left(B \geq \frac{\delta}{d} \right) \leq \left(5/\delta\right)^{4d} \cdot 2 \exp (-c\delta^2 N).\]
Note that for any self-adjoint operators $a,b \in \mathcal{B}(\C^d)$
\begin{equation} \label{eq} \left| \frac{1}{n} \sum_{i=1}^n \tr (U_i a U_i^\dagger b) \right| \leq \|a\|_1\|b\|_1 \left(A+\frac{1}{d} \right).\end{equation}
By a convexity argument, the supremum in $A$ can be restricted to pure states. Let $\phi,\psi$ be pure states and $\phi_0,\psi_0$ in $\mathcal{M}$ so that $\| \phi-\phi_0\|_1 \leq \delta, \|\psi-\psi_0\|_1 \leq \delta$.
Then 
\begin{eqnarray*} \left| \frac{1}{n} \sum_{i=1}^n \tr (U_i \phi U_i^\dagger \psi) - \frac{1}{d} \right| 
&\leq& \left| \frac{1}{n} \sum_{i=1}^n \tr (U_i \phi_0 U_i^\dagger \psi_0) - \frac{1}{d} \right| \\
& &+ \left| \frac{1}{n} \sum_{i=1}^n \tr (U_i (\phi-\phi_0) U_i^\dagger \psi_0)\right| + \left|\frac{1}{n}\sum_{i=1}^n \tr (U_i \phi U_i^\dagger (\psi-\psi_0) )\right| .
\end{eqnarray*}
Taking the supremum over $\phi,\psi$ and using twice \eqref{eq}, we get $A \leq B + 2 \delta (A + 1/d)$, and so
\[ A \leq \frac{1}{1-2\delta} \left(B + \frac{2\delta}{d} \right). \]
Choosing $\delta = \e/(3+2\e) \geq \e/5$ gives
\[ \P \left(A \geq \frac{\e}{d} \right) \leq \P \left(B \geq \frac{\delta}{d} \right) \leq 2 \left(\frac{25}{\e}\right)^{4d} \exp(-c\e^2 N/25). \] 
The last quantity is less than 1 provided $N \geq Cd/\e^2 \cdot \log(1/\e)$ for some constant $C$.

\begin{rk}
One checks (using the value $c=(6 \ln 2)^{-1}$ from \cite{hlsw}) that for $d$ large enough, the constant in our theorem can the chosen to $C = 150$. This is presumably far from optimal.
\end{rk}

\medskip

\noi Address : Université de Lyon, Université de Lyon 1, \\
CNRS, UMR 5208 Institut Camille Jordan, \\
Batiment du Doyen Jean Braconnier, \\
43, boulevard du 11 novembre 1918, \\
F - 69622 Villeurbanne Cedex, France.\\
\\
e-mail: \verb!aubrun@math.univ-lyon1.fr!

\end{document}